\documentclass[epj]{svjour}
\usepackage{graphics}

\def\schpt{S\raise0.4ex\hbox{$\chi$}PT}

\def\ltwid{{\,\raise.3ex\hbox{$<$\kern-.75em\lower1ex\hbox{$\sim$}}\,}}
\def\MeV{{\rm Me\!V}}
\def\chpt{\raise0.4ex\hbox{$\chi$}PT}
\def\msbar{{\overline{\rm MS}}}
\def\GeV{{\rm Ge\!V}}

\begin{document}
\title{2+1 flavor simulations of QCD with improved staggered quarks}
\author{Urs M. Heller
\thanks{\emph{Presented on behalf of the MILC Collaboration} }
}                     
\institute{American Physical Society, One Research Road, Ridge, NY 11961-9000,
USA
}
\date{Oct. 5, 2006}
%
\abstract{
The MILC collaboration has been performing realistic simulations of full QCD
with 2+1 flavors of improved staggered quarks. Our simulations allow for
controlled continuum and chiral extrapolations. I present results for the
light pseudoscalar sector: masses and decay constants, quark masses and
Gasser-Leutwyler low-energy constants. In addition I will present some
results for heavy-light mesons, decay constants and semileptonic form factors,
obtained in collaboration with the HPQCD and Fermilab lattice collaborations.
Such calculations will help in the extraction of CKM matrix elements
from experimental measurements.
} 
\maketitle

\section{Introduction}

QCD simulations with the effects of three light quark flavors fully
included have been a long standing goal of lattice gauge theorists.
The MILC collaboration has made significant advances towards this goal,
employing an improved staggered formalism, ``asqtad'' fermions
\cite{ASQTAD}. This formalism, and the ever increasing computational
resources available today, have enabled simulations with the strange
quark mass near its physical value, and the up and down quarks, taken
to be of equal mass, as light as $1/10$ the strange quark mass,
corresponding to pion masses as low as $240\; \MeV$. To
control the extrapolation to the continuum limit, simulations have
been done, and are still ongoing, at multiple lattice spacings,
$a \approx 0.15$, $0.12$, $0.09$ and $0.06$ fm. These are referred to
as coarser, coarse, fine and super-fine lattices, respectively.

All the lattices generated by the MILC collaboration are made available
to other researchers at the ``NERSC Gauge Connection''. They have been
widely used, for example by the LHPC and NPLQCD collaborations for the
study of nucleon properties, form factors and scattering lengths,
and by the Fermilab lattice, HPQCD and UKQCD collaborations for the study of
heavy quarkonia and heavy--light mesons. The last three joined efforts
with the MILC collaboration to validate these QCD simulations. They
compared selected, ``gold-plated'', {\it i.e.} well controlled,
quantities computed on the lattice with their experimentally well known
values. Agreement within errors of 1--3\% was found~\cite{PRL}.

Such a validation is, of course, valuable for any lattice simulation.
It is, in particular, needed for simulations with staggered fermions,
which rely on the so-called ``fourth-root trick'' to eliminate the extra
species of fermions, called ``tastes'', present with staggered fermions,
and thus to generate the correct number of sea quark flavors. A nice
review of the validity or possible problems with this fourth-root trick
can be found in~\cite{Sharpe_lat06}.

While the MILC collaborations has made simulations with lighter up and
down quark masses, $m_l$, than have been reached by other groups,
simulations at the physical light quark mass are still too costly,
even with the improved ``asqtad'' quarks. Therefore, extrapolations
in the light quark mass -- chiral extrapolations -- are still needed
to reach the physical value. Such extrapolations are well understood,
theoretically, based on chiral perturbation theory, \chpt. For results
from lattice QCD, it is important to include a treatment of lattice
effects into the chiral extrapolations. This holds, in particular,
for simulation results with staggered fermions, because of the taste
symmetry breaking effects. For this purpose, \chpt\ was adapted to
staggered quarks in staggered chiral perturbation theory,
\schpt~\cite{CB_SCPT}.

MILC's first set of accurate results for light pseudoscalars, based on
two lattice spacings (coarse and fine, {\it i.e.} $0.12$ and $0.09$ fm),
one simulation strange quark mass $m_s'$ and several simulation light
quark masses $m_l'$\footnote{We denote simulation masses by a prime, $m'$},
with the \schpt\ formalism for joint chiral and
continuum extrapolations, was published in 2004~\cite{PSEUDO04}.
To obtain quark masses, mass renormalization constants are needed.
The values with one-loop $Z$-factors, computed by members of the HPQCD
and UKQCD collaborations, appeared in~\cite{QMASS04}.

Since 2004, our simulations expanded in several ways. On the coarse,
$a=0.12$ fm, lattice a second simulation strange quark mass
$m_s'$ has been used, allowing interpolation to the physical strange
quark mass. On the fine, $a=0.09$ fm, lattice a simulation with a lighter
light quark mass, $m_l' \simeq 0.1 m_s$, has been done. A coarser
lattice ensemble, with $a=0.15$ fm, has been added, increasing the lever
arm for continuum extrapolations. And, finally, a super-fine lattice
ensemble, with $a=0.06$ fm, has been started. A run with $m_l' = 0.4 m_s'$,
corresponding to $m_\pi\!\approx\! 430\; \MeV$ is half-finished, and
lighter-mass simulations have begun.

\section{The light pseudoscalar sector}

\begin{figure}
\resizebox{0.45\textwidth}{!}{
\includegraphics{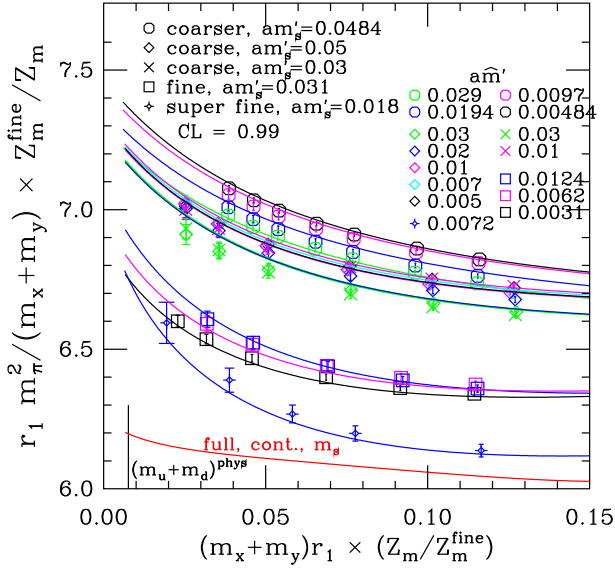}}
\caption{\label{MPIsq} The square of the pion mass divided by the sum of
the valence quark masses as a function of the sum of the valence
quark masses in units of $r_1$.}
\end{figure}

\begin{figure}
\resizebox{0.45\textwidth}{!}{
\includegraphics{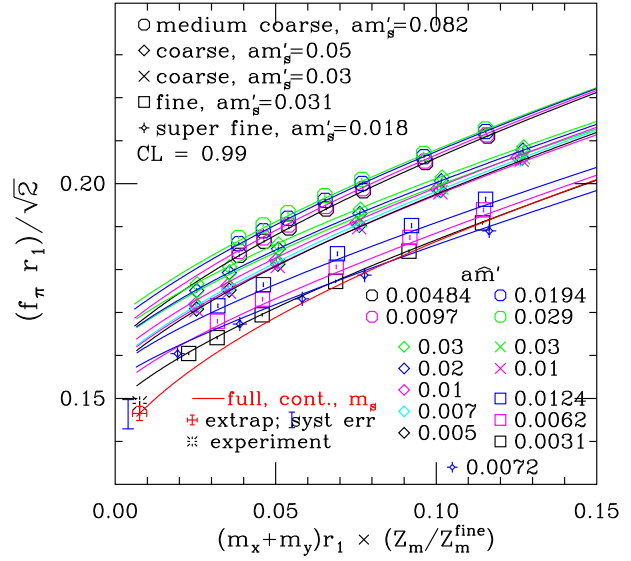}}
\caption{\label{FPI} The pion decay constant as a function of the sum of
the valence quark masses in units of $r_1$.}
\end{figure}

The \schpt\ fitting is illustrated in Figs.~\ref{MPIsq} and \ref{FPI}.
Dimensionful quantities are given in units of $r_1 = 0.318(7)$~fm,
a scale related to the heavy quark potential~\cite{MILC-POTENTIAL}.
The quark masses are renormalized (at one-loop) relative to those on
the fine lattice. The data is accurate enough and at small enough
quark masses so that the effects of chiral logs are evident.
The red lines are the fit functions in ``full continuum QCD'' (valence
and sea quark masses set equal) after extrapolation of parameters to the
continuum limit.

Preliminary numerical results for $f_\pi$ and $f_K$ obtained from our most
recent data are\footnote{The numbers here are updated as of
Lattice 2006~\cite{FPILAT06}.}
\begin{eqnarray}\label{eq:f_results}
f_\pi  =  128.6 \pm 0.4\pm 3.0 \; \MeV && [{129.5 \pm 0.9\pm 3.5 \; \MeV}] \\
f_K  =    155.3 \pm 0.4\pm 3.1 \; \MeV &&  [{156.6 \pm 1.0\pm 3.6 \; \MeV}] \\
f_K/f_\pi   = 1.208(2)({}^{+\phantom{1}7}_{-14}) &&   [{ 1.210(4)(13)}].
\end{eqnarray}
Here the numbers on the left are the new values, and those on the
right in square brackets are from Ref.~\cite{PSEUDO04}.
In each case the first error is statistical, and the second systematic.

The lattice QCD value of $f_K/f_\pi$ can be combined with experimental
data to extract the CKM matrix element $V_{us}$~\cite{MARCIANO}.
We obtain $|V_{us}|=0.2223({}^{+26}_{-14})$, with an accuracy
comparable to, and a value compatible with, the latest (2006) PDG value,
$V_{us}\!=\!0.2257(21)$~\cite{PDG}.

The up, down and strange quark masses can be determined from the
masses of the $\pi$ and $K$ mesons using the \schpt\ fits.
To relate these masses to the experimental ones requires some continuum
input for isospin breaking and electromagnetic effects~\cite{CONT}.
Details can be found in Ref.~\cite{PSEUDO04}.
Preliminary results from our current data set, evaluated at scale
$\mu \!= \! 2\, \GeV$, are:
\begin{eqnarray}
m_s^\msbar = 90(0)(5)(4)(0)\;\MeV &&  [76(0)(3)(7)(0)\;\MeV]\\
m_u^\msbar = 2.0(0)(1)(1)(1)\;\MeV &&  [1.7(0)(1)(2)(2)\;\MeV] \\
m_d^\msbar = 4.6(0)(2)(2)(1)\;\MeV  &&   [3.9(0)(1)(4)(2)\;\MeV] \\
m_s/m_l = 27.2(0)(4)(0)(0) &&  [27.4(1)(4)(0)(1)] \\
m_u/m_d = 0.42(0)(1)(0)(4)  && [0.43(0)(1)(0)(8)] ~.
\end{eqnarray}
Again, new results are on the left, and earlier ones from
Refs.~\cite{PSEUDO04,QMASS04} are in square brackets on the right.
Errors are from statistics, simulation systematics, perturbation theory
and electromagnetic effects.
The main difference between the new and old results comes from the use
of a two-loop mass renormalization constant~\cite{MTH}, compared to
the one-loop one used previously. A non-perturbative mass renormalization
calculation is in progress.
For further results, including Gasser-Leutwyler low-energy constants,
see Ref.~\cite{FPILAT06}.

\section{Heavy-light meson physics}

\begin{figure}
\resizebox{0.45\textwidth}{!}{
\includegraphics{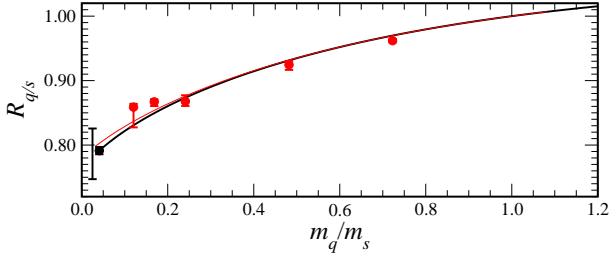}}
\caption{\label{R_q} The ratio $R_{q/s} = f_D \sqrt{m_D} / f_{D_s}
\sqrt{m_{D_s}}$ as function of $m_q/m_s$ together with the \schpt\ fit.
The black line and extrapolated point are obtained after removing
${\cal O}(a^2)$ effects from the fit, representing a continuum
estimate.}
\end{figure}

In a joint effort with the Fermilab lattice and HPQCD collaborations,
we used our full QCD ensembles to study properties of heavy-light
mesons. For the heavy $c$ and $b$ quarks we used clover fermions with
the Fermilab interpretation~\cite{Fermilab}, while ``asqtad''
fermions where used for the light valence quarks. This allows to go
much closer to the physical light quark mass than was achieved previously.
Use of \schpt, adopted to heavy-light mesons~\cite{HL-SCHPT}, makes the
necessary remaining chiral extrapolation well controlled, as illustrated
in Fig.~\ref{R_q}.

We found the lattice predictions \cite{FD_PRL,FD_FBLAT06}
\begin{eqnarray}
f_{D_s} & = &  249 \pm 3 \pm 16 \; \MeV\ , \\
f_D & = &  201 \pm 3 \pm 17 \; \MeV\ , \\
f_{D_s} / f_D  & = & 1.21 \pm .01 \pm .04 \ ,
\end{eqnarray}
compared to the later experimentally measured values from leptonic decays
\cite{CLEO_fD,BABAR_fDs}
\begin{eqnarray}
f_{D_s} & = &  283 \pm 17 \pm 7 \pm 14 \; \MeV\ , \\
f_{D^+} & = &  222.6 \pm 16.7 {}^{+2.8}_{-3.4}\; \MeV\ . 
\end{eqnarray}

With the Fermilab lattice and HPQCD collaborations, we are also computing
form factors for semileptonic $D \to \pi/K$ and $B \to \pi/D$ decays.
The heavy-to-light decay amplitudes are parametrized as
\begin{equation}
\langle P | V^\mu | H \rangle = f_+(q^2) (p_H + p_P - \Delta)^\mu
 + f_0(q^2) \Delta^\mu \ ,
\end{equation}
where $\Delta^\mu = (m^2_H - m^2_P) q^\mu / q^2$.
We can compare our calculation of $f^{(K)}_+(q^2)$ for
$D^0 \to K^- l^+ \nu$~\cite{Semilept} with the recent measurement
by Belle~\cite{Belle_f+}.

\begin{figure}
\resizebox{0.45\textwidth}{!}{
\includegraphics{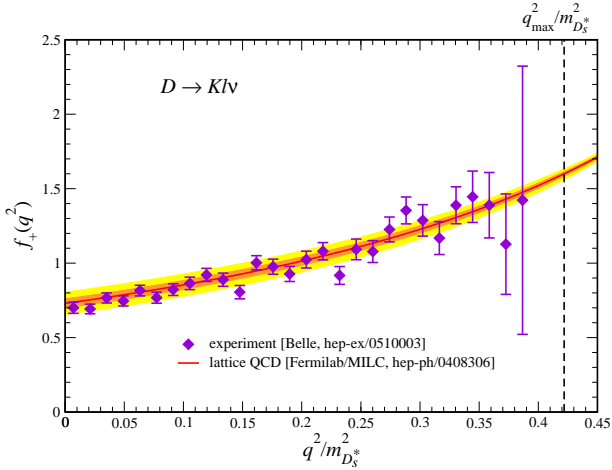}}
\caption{\label{f+} The form factor $f^{(K)}_+(q^2)$ for $D^0 \to
K^- l^+ \nu$ compared to recent measurements by Belle~\cite{Belle_f+}.}
\end{figure}

\section{Conclusions}

The MILC collaboration's QCD simulations with $2+1$ flavors have been
designed to have all errors, in particular from chiral and continuum
extrapolations, well controlled. For many quantities they have reached
hitherto unprecedented accuracy, so that the lattice QCD calculations
are starting to have an impact on current experimental physics programs.
For example, using only our lattice calculation for $f_K/f_\pi$ as
a theoretical input, the CKM matrix $V_{us}$ can be determined with
an accuracy comparable to the latest PDG result. A determination
of the entire CKM matrix with lattice QCD results based on the MILC
ensembles of gauge configurations as the only theoretical input
has been given in Ref.~\cite{OKAMOTO}.


\end{document}